\documentclass{aipproc}

\layoutstyle{6x9}

\SetInternalRegister\hbadness{8000} 

\newcommand\doingARLO[2][]{%
  \ifx\mmref\undefined #1\else #2\fi }

\def \be{\begin{equation}}
\def \ee{\end{equation}}
\def \berr{\begin{eqnarray}}
\def \err{\end{eqnarray}}
\def \nn{\nonumber}

\def \d{\delta}

\def \L{\Lambda}

\def \eps{\varepsilon}

\def \p{\varphi}

\def \A{{\cal A}}
\def \D{{\cal D}}
\def \F{{\cal F}}

\def \S{{\cal S}^2}

\def \RR{{\cal R}}
\def \TR{\tilde{\cal R}}
\def \NN{{\cal N}}

\def \Tr{\mbox{Tr}}

\def \Fun{{\rm Fun}}

\def \({\left(}
\def \){\right)}
\def \<{\langle}
\def \>{\rangle}
\def \[{\left[}
\def \]{\right]}

\def\tens{\mathop{\otimes}}

\def\id{\rm id}
\newcommand{\tr}{\triangleright}
\newcommand{\ttr}{\tilde\triangleright}

\newcommand \one{{\bf 1}}

\def\reps{representations }
\def\rep{representation }

\newcommand \reals{I \! \! R}
\newcommand \compl{C \! \! \! \! {\scriptscriptstyle {}^{{}_|}}\ }
\def\R{{\reals}}
\def\C{{\compl}}

\renewcommand{\theequation}{\thesection.\arabic{equation}}

\renewcommand{\theequation}{\thesection.\arabic{equation}}

\begin{document}

\title{Quantum Field Theory on the $q$--deformed Fuzzy Sphere}


\author{H. Steinacker}{
   address={Laboratoire de Physique Th\'eorique et Hautes Energies\\
        Universit\'e de Paris-Sud, B\^atiment 211, F-91405 Orsay and\\[2ex]
        Sektion Physik der Ludwig--Maximilians--Universit\"at\\
        Theresienstr.\ 37, D-80333 M\"unchen  \\[1ex]}, 
  email={Harold.Steinacker@th.u-psud.fr},
  thanks={This work was supported in part by the DFG fellowship STE 995/1-1}
}

\copyrightyear{2001}

\begin{abstract}
We discuss the second quantization of scalar field theory on the 
$q$--deformed fuzzy sphere $S^2_{q,N}$ for $q \in \R$, using a 
path--integral approach. We find quantum field theories which are
manifestly covariant under $U_q(su(2))$, have a smooth limit $q \rightarrow 1$,
and satisfy positivity and twisted bosonic symmetry properties.
Using a Drinfeld twist, they are equivalent to ordinary but
slightly ``nonlocal'' QFT's on the undeformed fuzzy sphere, 
which are covariant under $SU(2)$.

\end{abstract}

\date{\today}

\maketitle

\section{Introduction}

In this paper, we first give a short introduction to the $q$--deformed
fuzzy sphere, and then discuss some aspects of second quantization 
on this space. This is essentially a short introduction to the
more extensive discussion in \cite{hste-ours_2}.
Much of the considerations concerning the second quantization 
generalize to other, higher--dimensional  
$q$--deformed spaces.

The $q$--deformed fuzzy sphere 
$S^2_{q,N}$ is a $q$--deformed version of the ``ordinary'' fuzzy sphere
$S_N^2$ \cite{hste-madore}.
The algebra of functions on $S^2_{q,N}$ is isomorphic to 
the matrix algebra $Mat(N+1,\C)$, but viewed as a $U_q(su(2))$--module
algebra. It admits additional structure compatible with 
covariance under the Drinfeld--Jimbo quantum group
$U_q(su(2))$, such as an invariant integral and a differential calculus.
It can be defined for both $q \in \R$ and $|q| = 1$, however 
we restrict ourselves to the case 
$q \in \R$ here. Then $S^2_{q,N}$ is precisely the ``discrete series'' of 
Podles spheres \cite{hste-podles}. Moreover, we only consider 
scalar fields for simplicity.
A much more detailed description of $S^2_{q,N}$ has been given in 
\cite{hste-ours_1}. This space is of interest in the context of 
$D$--branes on the $SU_k(2)$ WZW model, 
as discussed by  Alekseev, Recknagel and Schomerus \cite{hste-ars}. 
These authors extract an ``effective'' algebra of functions on the
$D$--branes from the OPE of the boundary vertex operators, which 
is twist--equivalent \cite{hste-ours_1}
to the space of functions on $S^2_{q,N}$ for $q$ a root of unity.

\section{The space $S_{q,N}^2$}
 
Consider the spin $\frac N2$ \rep of $U_q(su(2))$, 
$$
\rho: U_q(su(2)) \rightarrow Mat(N+1,\C), 
$$
which acts on $\C^{N+1}$. It can be used to define the
quantum adjoint action of $U_q(su(2))$ on the set of matrices $Mat(N+1,\C)$, 
by 
$$
u \tr_q M = \rho(u_1) M \rho(Su_2).
$$
The usual matrix algebra $Mat(N+1,\C)$ thereby becomes a 
$U_q(su(2))$--module algebra, which means that 
$u \tr_q (ab) = (u_{(1)} \tr_q a) (u_{(2)} \tr_q b)$
for $a, b \in Mat(N+1,\C)$. Here $\Delta(u) = u_{(1)} \tens u_{(2)}$ 
denotes the coproduct of $u \in U_q(su(2))$. 
$S_{q,N}^2$ is defined to be precisely this $U_q(su(2))$--module algebra
$Mat(N+1,\C)$,
together with some additional structure. It is easy to see that 
under the (adjoint) action of $U_q(su(2))$, 
it decomposes into the irreducible \reps 
\be
S^2_{q,N}= Mat(N+1,\C) = (1) \oplus (3) \oplus ... \oplus (2N+1),
\label{hste-decomp}
\ee
where $(2K+1)$ is the spin $K$ \rep of $U_q(su(2))$.
This is the analog of the decomposition of functions on the sphere into
spherical harmonics, which it is truncated on the fuzzy spheres.
Let $\{x_i\}_{i = +,-,0}$ be the weight basis of the spin 1 components
in (\ref{hste-decomp}),
so that $u \tr_q x_i = x_j \pi^j_i(u)$ for $u \in U_q(su(2))$.
One can show that they satisfy the relations 
\berr
\eps^{ij}_k x_i x_j &=& \L_{N} x_k,  \nn\\
g^{ij} x_i x_j  &=& R^2. \nn
\err
Here 
$$
\L_{N} = R\; \frac{[2]_{q^{N+1}}}{\sqrt{[N]_q [N+2]_q}},
$$
$[n]_q = \frac {q^n-q^{-n}}{q-q^{-1}}$, and 
$\eps^{ij}_k$ and $g^{ij}$ are the $q$--deformed 
invariant tensors. For example, $\eps^{33}_3 = q^{-1} - q$, and 
$g^{1 -1} = -q^{-1}, \;\; g^{0 0} = 1, \;\; g^{-1 1} = -q$. 
In \cite{hste-ours_1}, these relations 
were derived using a Jordan--Wigner construction. 
For $q=1$, the relations of $S^2_N$ are recovered.

\paragraph{Integration}
The unique invariant integral of a function $f \in S^2_{q,N}$ is 
given by its quantum trace over $Mat(N+1,\C)$,
$$
\int f :=  \frac {4\pi R^2}{[N+1]_q} \Tr_q(f) = 
           \frac {4\pi R^2}{[N+1]_q} \Tr(f \;q^{-H}),
$$
normalized such that $\int 1 =4\pi R^2$.
Here $H$ is the Cartan generator of $U_q(su(2))$.
Invariance means that $\int u \tr_q f = \eps(u) \; \int f$.

\paragraph{Real structure}
In order to define a {\em real} noncommutative space, one must 
specify a star structure on the algebra of functions. 
The star of an element $f$ 
is simply defined to be the hermitean adjoint of the matrix 
$f\in S^2_{q,N} = Mat(N+1,\C)$. 
In terms of the generators  $x_i$, this  becomes 
$$
x_i^* = g^{ij} x_j,
$$ 
since $q$ is real.

$S^2_{q,N}$ admits additional structure, in particular a differential calculus.
While the calculus is very interesting in the context of gauge theories, 
we shall not discuss it here.
The interested reader is referred to \cite{hste-ours_1}.
However, we do need a Laplacian 
in order to write down Lagrangians and actions. 
While it can naturally be defined using the differential 
calculus as $\Delta = \ast_H d \ast_H d$, 
we give an ad--hoc definition here for simplicity. 
Assume that $\{\psi_{K,n}(x)\}_{K,n} \subset S^2_{q,N}$
is a weight basis of the spin $K$ \rep of $U_q(su(2))$, so that 
\be
u \tr_q \psi_{K,n}(x) = \psi_{K,m}(x) \pi^{m}_{n}(u).
\label{hste-psi_covar}
\ee
It can be normalized such that 
$$
\int \psi_{K,n}(x) \psi_{K',m}(x)
= \d_{K,K'} \; g^K_{n,m}.
$$
The Laplacian is then given by
$$
\Delta \psi_{K,n}(x)\; = 
     \frac 1{R^2}\; [K]_q [K+1]_q \;\psi_{K,n}(x).
$$

\section{Scalar field theory on $\S_{q,N}$}

We can now write down Lagrangians and actions defining scalar field 
theory on $\S_{q,N}$. Consider for example
$$
S[\Psi] = - \int\frac 12\;\Psi \Delta \Psi + \lambda\Psi^4
 = S_{free}[\psi] + S_{int}[\psi]
$$
where 
\be
\Psi(x) = \sum_{K,n} \psi_{K,n}(x)\;  a^{K,n}.
\label{hste-psi_field}
\ee
The free action can be rewritten as\footnote{the tilde labels objects 
associated with 
$\tilde U_q(su(2))$, which is another copy of $U_q(su(2))$ but with 
reversed coproduct, see \cite{hste-ours_2}.} 
$$
S_{free}[\Psi] = - \sum_{K,n} \frac 12 D_K \;\tilde g^K_{nm}
               a^{K,m} a^{K,n}  
$$
In general, actions will be
polynomials in the variables $a^{K,n}$ which are invariant under 
$\tilde U_q(su(2))$.

We want to discuss the second quantization of such models,
as in \cite{hste-ours_2}. 
On the undeformed fuzzy sphere, this is fairly straightforward 
\cite{hste-grosse,hste-madore}:
the coefficients $a^{K,n}$ are considered as complex numbers
or more precisely as coordinate functions\footnote{with star 
structure $a_i^* = g_c^{ij} a_j$}
on the representation space $\R^{2K+1}$,
so that the actions can be considered as polynomials in the algebra
$\A = \tens_{K=0}^{N} Fun(\R^{2K+1})$.
The ``path integral'' is then simply 
the product of the ordinary integrals over the coefficients $a^{K,n}$,
i.e. over $\prod_K \R^{2K+1}$.
This defines a quantum field theory which has a $SO(3)$ rotation symmetry, 
because the path integral is invariant.

In the $q$--deformed case, this is not as easy, and needs some discussion.
We certainly want the models to have a $U_q(su(2))$ symmetry at the quantum 
level. This means that the coefficients $a^{K,n}$ in 
(\ref{hste-psi_field}) 
must be considered as \reps of $U_q(su(2))$. 
In order to be able to do calculations, we also require
that the $a^{K,n}$ generate some kind of 
algebra $\A$; this is almost a tautology. 
This strongly suggests that $\A$ should be a $U_q(su(2))$--module algebra.
We do not have in mind here an algebra of field operators, 
which in fact would not be appropriate in the Euclidean case even for $q=1$. 
Rather, $\A$ should be an analog of the algebra 
of coordinate functions on configuration space as above
for $q=1$, i.e. some deformed version of $\tens_{K=0}^{N} Fun(\R^{2K+1})$.
Our goal is
to define correlation functions of the fields (\ref{hste-psi_field}),
which after ``Fourier transform'' amounts to defining
\be
  \langle a^{K_1,n_1} a^{K_2,n_2} ... a^{K_k,n_k}\rangle =: 
   \langle P(a) \rangle \;\; \in \C,
\label{hste-correlation-function}
\ee 
perhaps by some kind of a path integral
$ \langle P(a) \rangle = \frac 1{\NN} \int \D a\; e^{-S[\Psi]}  P(a)$.
$P(a)$ will denote some polynomial in the variables $a^{K,n}$ from now on.

It follows immediately from these considerations that $\A$ cannot be
commutative, because the coproduct of $U_q(su(2))$ is not cocommutative. 
In particular, the $a^{K,n}$ cannot be ordinary complex numbers. 
Therefore an ordinary integral over commutative modes $a^{K,n}$ would 
violate $U_q(su(2))$ invariance at the quantum level. 
In some sense, this means that on $q$--deformed spaces, a second quantization
is required by consistency.
There is one more essential requirement: $\A$ should have
the same Poincar\'e series as classically, 
i.e. the dimension of the space of polynomials at a given degree should be
the same as in the undeformed case. 
This is in fact precisely the content of a symmetrization postulate, and
it is of course an essential physical requirement at least for 
low energies, in order to have the correct number of degrees of freedom.
It means that the ``amount of information'' contained in the 
$n$--point functions should be the same as for $q=1$,
so that a smooth limit $q \rightarrow 1$ is conceivable. 
In other words, we want to consider ordinary bosons\footnote{we do not 
consider fermions here}.
While some proposals have been given in the literature
\cite{hste-oeckl} how to define 
QFT on spaces with quantum group symmetry, none of them seems to
satisfy these requirements.

On a more formal level, we impose the following requirements 
\cite{hste-ours_2}:
\begin{itemize}

\item[(1)] {\em Covariance:}
$$
\langle u\; \ttr_q P(a) \rangle = \eps_q(u) \; \langle P(a) \rangle,
$$
which means that the $\langle P(a) \rangle$ are invariant 
tensors of $\tilde U_q(su(2))$,

\item[(2)] {\em Hermiticity:} 
$$
\langle P(a) \rangle^* = \langle P^*(a) \rangle
$$
for a suitable involution $*$ on $\A$,

\item[(3)] {\em Positivity:}
$$
\langle P(a)^* P(a)\rangle \geq 0,
$$

\item[(4)] {\em Symmetry} 

under permutations of the fields, by which we mean that
the polynomials in the $a^{K,n}$ can be ordered as usual, 
i.e. the Poincar\'e series of 
$\A$ should be underformed.
\end{itemize}

A slight refinement will be needed later.

Unfortunately, there is no obvious candidate for an associative 
$U_q(su(2))$--module algebra $\A$ with the same 
Poincar\'e series as $\tens_{K=0}^{N} Fun(\R^{2K+1})$
(except for small $N$).
We will therefore construct a suitable quasiassociative algebra $\A$ 
which is a star--deformation of the commutative 
$\tens_{K=0}^{N} Fun(\R^{2K+1})$.
We want to emphasise that quasiassociativity
is in no way inconsistent with the 
usual axioms of quantum mechanics, because the algebra $\A$ will 
not be interpreted as algebra of observables;
it is only a tool which is useful to calculate correlation functions, 
just like Grassman variables are used to calculate fermionic
correlation functions. In fact, it is possible to avoid
the use of quasiassociative algebras alltogether, see \cite{hste-ours_2}.
Any lingering doubts can be eliminated by showing the equivalence of 
our models to ordinary QFT on the undeformed 
fuzzzy sphere, with slightly derformed interactions.

The chosen approach is rather general and is applicable 
in a more general context, such as for
higher--dimensional theories.

\subsection{The quasiassociative star product}

As discussed, we assume that the coefficients 
$a^{K,n}$ transform in the spin $K$ \rep of
$\tilde U_q(su(2))$,
\be 
u \; \ttr_q a^{K,n} = \pi^{n}_{m}(\tilde S u) \; a^{K,m}.
\label{hste-a_covar}
\ee  
Let $\p$ be the algebra (not coalgebra!)--isomorphism \cite{hste-drinfeld} 
$$
\varphi: \tilde U_q(su(2)) \rightarrow U(su(2))[[h]],
$$
where $q = e^h$. Moreover, let
$\F = \F_1\tens \F_2 \in U(su(2))[[h]]\tens U(su(2))[[h]]$
be the ``Drinfeld--twist'' \cite{hste-drinfeld} 
which relates the Hopf algebras 
$\tilde U_q(su(2))$ and $U(su(2))$,
and satisfies among others
\berr
\F &=& \one\tens\one +o(h),  \nn\\
(\varepsilon\tens \id)\F &=& \one=(\id\tens \varepsilon)\F, 
                \label{hste-cond2}\\
(\p\tens\p) \tilde\Delta_q(u)&=& \F\Delta(\p(u))\F^{-1},  \nn\\
(\p\tens\p)\RR &=& \F_{21} q^{t\over 2}\F^{-1}.        \nn
\err
for any $u \in \tilde U_q(su(2))$.
Using this twist, there is an action of $U(su(2))$ on the coefficients 
$a^{K,n}$, by $u \tr a^{K,n} = \p^{-1}(u)\;\ttr_q  a^{K,n}$.
Hence we can consider the usual commutative 
algebra $\A^K := Fun(\R^{2K+1})$ generated
by the $a^{K,n}$, and view it as a $U(su(2))$--module
algebra $(\A^K, \cdot, \tr)$. We now then define a new
multiplication on the same space $\A$ by
\be
a \star b := (\F^{-1}_1\tr a) \cdot(\F^{-1}_2\tr b) 
            = \cdot(\F^{-1} \tr(a\tens b))
\label{hste-star}
\ee 
for any $a,b \in \A^K$. This is analogous to the Moyal product in 
deformation quantization. 
It is easy to verify that it satisfies
$$
u\; \ttr_q (a \star b) = \star \( \tilde \Delta_q(u) \ttr_q a \tens b\),
$$
which means that $(\A^K, \star,\ttr_q)$ is a  $\tilde U_q(su(2))$--module 
algebra. It follows from (\ref{hste-cond2}) that if $a$ is invariant under 
$\tilde U_q(su(2))$, then it is also central in $(\A^K, \star)$.
Moreover, the following commutation relations 
are derived in \cite{hste-ours_1}:
\be
a_i \star a_j - a_k \star a_l \; \TR_{ij}^{lk} = 0,
\label{hste-CR}
\ee
were $\TR_{ij}^{lk}$ is obtained from the universal $\RR$ matrix of 
$\tilde U_q(su(2))$.
This new product is not associative, but quasiassocative:
$$
(a\star b)\star c = (\tilde\phi_1\tr a)\star\((\tilde\phi_2\tr b)
         \star(\tilde\phi_3\tr c)\).
$$
Here 
$$
\tilde\phi := (\one \tens \F)
   [(\id \tens \Delta)\F ][(\Delta\tens \id)\F^{-1}](\F^{-1}\tens \one) 
$$
is the coassociator, which is invariant under $U_q(su(2))$ and
closely related to the KZ equation \cite{hste-drinfeld}.
It is much easier to work with than
the Drinfeld-twist $\F$, which in fact is never needed 
explicitly. 

Finally, $(\A, \star,\ttr_q)$ is defined as in (\ref{hste-star}), 
applied to any element of $\A = \tens_K \Fun(\R^{2K+1})$.
Polynomials $P_{\star}(a)$ must now be given including some
``bracketing''. Nevertheless, the Poincar\'e series of $\A$ 
is undeformed, because the vector space $\A$ is undeformed, and 
the new product preserves the grading. Different bracketings can always be 
related using the coassociator.

Invariant actions are now considered of the form 
\be
S_{int}[\Psi] = \int \Psi(x) \star (\Psi(x) \star \Psi(x))
= I^{(3)}_{K,K',K'';\; n,m,l} \; a^{K,n}\star (a^{K',m}\star a^{K'',l}),
\label{hste-example}
\ee
which are invariant polynomials in $\A$.

\subsection{Quantization}

The path integral should be a ``functional'' on 
$\A$ which is invariant under $\tilde U_q(su(2))$.
As in deformation quantization, we
view $\A^K$ as the vector space of complex--valued functions on 
$\R^{2K+1}$, and consider the usual classical integral over $\R^{2K+1}$.
Observe that it is also invariant under the action $\ttr_q$ 
(\ref{hste-a_covar}) of $\tilde U_q(su(2))$, because 
the algebra structure does not enter here at all.
Explicitly, let $\int d^{2K+1}\!a^{K}\; f$ be the ordinary integral 
of $f \in \A^K$ over $\R^{2K+1}$.
The path integral is then defined as
$$
\int \D \Psi \; f[\Psi] := \int \prod_K d^{2K+1}\!a^{K}\; f[\Psi],
$$
where $f[\Psi] \in \A$ denotes any integrable function (in the usual sense)
of the variables $a^{K,m}$. It is by construction 
invariant under $\tilde U_q(su(2))$.

Correlation functions can now be defined as 
functionals of ``bracketed polynomials'' 
$P_{\star}(a) = a^{K_1,n_1} \star (a^{K_2,n_2} \star (...\star a^{K_l,n_l}))$
in the  field coefficients by
\be
\langle P_{\star}(a)\rangle := \frac{\int  \D \Psi\; e^{-S[\Psi]} P_{\star}(a)}
                                    {\int  \D \Psi\; e^{-S[\Psi]}}.
\label{hste-correlation}
\ee
This is natural, because all invariant actions $S[\Psi]$
commute with the generators $a^{K,n}$.
Strictly speaking there should be a factor $\frac 1{\hbar}$ in front
of the action, which we shall omit.
Invariance of the action $S[\Psi] \in \A$ implies that
$$
\langle u\; \ttr_q P_{\star}(a) \rangle  = 
             \eps_q(u)\; \langle P_{\star}(a) \rangle.
$$
By construction, the number of
independent modes of a polynomial $P_{\star}(a)$ with given degree
is the same as for $q=1$. One can in fact order them, using
quasiassociativity together with the commutation relations
(\ref{hste-CR}).
Therefore the symmetry requirement (4) above is satisfied.
Using a suitable formalism, one can show that the 
requirements (2) and (3) are satisfied as well, 
see \cite{hste-ours_2}.

The field theories defined in this way are equivalent to ordinary 
QFT's on the undeformed fuzzy sphere, with slightly nonlocal interactions. 
Consider an interaction term of the form (\ref{hste-example}).
If we write down explicitly the definition of the $\star$ product
of the $a^{K,n}$ variables, then it can be viewed as an interaction
term of $a^{K,n}$ variables with a tensor which is invariant under 
the {\em undeformed} $U(su(2))$, obtained from the 
$\tilde U_q(su(2))$--invariant tensor by 
multiplication with the twist $\F = \one + o(h)$.
In other words, the above actions can also be viewed
as actions on the undeformed fuzzy sphere $S^2_{q=1,N}$, with interactions
which are slightly ``nonlocal'' in the sense of  $S^2_{q=1,N}$, i.e.
they are given by traces of products of matrices only to the lowest 
order in $h$. Upon spelling out the $\star$ product in the 
correlation functions (\ref{hste-correlation}) as well, 
they can be considered as
ordinary correlation functions of a slightly nonlocal field theory
on $S^2_{q=1,N}$, disguised by the transformation $\F$.
Therefore $q$--deformation simply amounts to some kind of 
nonlocality of the interactions. 
A similar interpretation is well--known in the context of
field theories on spaces with a Moyal product.

Finally, it is possible to calculate correlators in perturbation 
theory, and to derive an analog of Wicks theorem. For lack of space, 
the reader is referred to \cite{hste-ours_2}.

\begin{theacknowledgments}

The author thanks J. Lukierski and the organizers of the XXXVII Winter School 
of Theoretical Physics in Karpacz for hospitality and stimulating atmosphere
during the school. It is also a pleasure to thank H.Grosse and J. Madore 
for collaboration on the two papers \cite{hste-ours_1,hste-ours_2} 
underlying this report,
as well as  G. Fiore, S. Majid, R. Oeckl and 
J. Pawelczyk for useful discussions.

\end{theacknowledgments}

\end{document}